\documentclass[letter]{aa}

\usepackage{txfonts}

\newcommand{\hi}{\mbox{H\,{\sc i}}}

\begin{document}

\title{Metals and dust in the neutral ISM:\\the Galaxy, Magellanic Clouds, and damped Lyman-$\alpha$ absorbers}

   \author{Annalisa De Cia\inst{1}}

\institute{
 European Southern Observatory, Karl-Schwarzschild Str. 2, 85748 Garching bei M\"unchen, Germany\\
 \email{adecia@eso.org}
 }

   \date{Received Month dd, 2017; accepted Month dd, 2017}

  \abstract
   {The presence of dust in the neutral interstellar medium (ISM) dramatically changes the metal abundances that we measure. Understanding the metal content in the neutral ISM, and a direct comparison between different environments, has been hampered to date because of the degeneracy to  the observed
ISM abundances caused by the effects of metallicity, the presence of dust, and nucleosynthesis.}
   {We   study the metal and dust content in the neutral ISM consistently in different environments, and assess  the universality of recently discovered sequences of relative abundances. We also intend to assess the validity of [Zn/Fe] as a tracer of dust in the ISM. This has recently been cast into doubt based on observations of stellar abundances, and needs to be addressed before we can safely use it to study the ISM.}
   {In this letter we present a simple comparison of relative abundances observed in the neutral ISM in the Galaxy, the Magellanic Clouds, and damped Lyman-$\alpha$ Absorbers (DLAs). The main novelty in this comparison is the inclusion of the Magellanic Clouds.}
   {The same sequences of relative abundances are valid for the Galaxy, Magellanic Clouds, and DLAs. These sequences are driven by the presence of dust in the ISM and seem `universal'.}
   {The metal and dust properties in the neutral ISM appear to follow a similar behaviour in different environments. This suggests that a dominant fraction of the dust budget is built up from grain growth in the ISM depending of the physical conditions and regardless of the star formation history of the system. In addition, the DLA gas behaves like the neutral ISM, at least from a chemical point of view. Finally, despite the deviations in [Zn/Fe] observed in stellar abundances, [Zn/Fe] is a robust dust tracer in the ISM of different environments, from the Galaxy to DLAs.}


   \maketitle
%
%
\section{Introduction}

In the early years of interstellar medium (ISM) UV studies, it was already  clear that refractory elements, i.e. those with high condensation temperatures as measured in the lab, show very low abundances in the Galaxy, down to a small percent of the abundances of more volatile metals. As an example, 90\% -- 99\% of the iron is missing from the gas-phase ISM in the Galaxy. The reason why large fractions of the refractory metals are missing from the gas phase is that they are instead incorporated into dust grains  \citep{Field74,Jenkins86,Cardelli93,Hobbs93,Phillips82,Phillips84,Welty95,Savage96}. This phenomenon is called dust depletion; its strength depends on the density of the gas, and it heavily affects the measurements of gas-phase metal abundances in the ISM, whether  in the Galaxy \citep[][and references above]{Jenkins09,DeCia16},  in the Magellanic Clouds \citep[e.g.][]{Tchernyshyov15,Jenkins17}, or in high-$z$ systems \citep{Pettini94, Kulkarni97,Vladilo98,Pettini00,Savaglio01,Ledoux02,Prochaska02,Vladilo02,Dessauges-Zavadsky06,Rodriguez06, Meiring06,Vladilo11,Som13,DeCia13,DeCia16,Quiret16,Wiseman17,DeCia18} such as damped Lyman-$\alpha$ absorbers \citep[DLAs, defined by $N(HI) \geq 2 \times 10^{20}$ cm$^{-2}$; e.g.][]{Wolfe86}.

By the neutral ISM we mean the \hi{} gas in galaxies, at any redshift, which is more extended than the stellar disk both in radius and height, is dominated by \hi{} and singly ionized metals, has relatively small velocities (within a few hundred km/s), and has a small ionization fraction \citep[typically up to 2\% in the warm neutral medium in the Galaxy; e.g.][]{Draine11}. For example, the extra-planar \hi{} gas in the Galaxy with $N(HI) \sim 10^{19}$--$10^{21}$ cm$^{-2}$ extends up to $\sim1.6$ kpc, far above the stellar disk \citep{Marasco11}. Local gas-rich dwarf galaxies, which may resemble part of the DLA population at higher $z$, are gas dominated (gas-to-baryonic fractions of 50 –- 90\%). The radii of the main \hi{} disks of both dwarf and spiral galaxies are more extended than their stellar disks by a factor of 4 on average \citep{Lelli16}.  

The understanding of metal content in the neutral ISM and a direct comparison of different environments has been hampered for years by the degeneracy to the metal
budget of the neutral ISM caused by the effects of metallicity, dust depletion, and nucleosynthesis.  A leap forward was made by \citet{DeCia16}, who found empirical relations (sequences) between relative abundances observed in the ISM of the Galaxy and DLAs. This was used  in turn  to develop a new method for characterizing dust depletion in different environments. 

In this letter we report a simple comparison of the observed relative abundances in the Galaxy, DLAs, and the Magellanic Clouds. We discuss the consequences of the potential universality of the sequences of relative abundances (Sect. \ref{sec ISM}). In addition, we discuss the reliability of [Zn/Fe] as a dust tracer in the ISM, and the differences between [Zn/Fe] measured in stellar abundances and in the ISM (Sect. \ref{sec znfe}). Conclusions are briefly drawn in Sect. \ref{sec conclusions}. Relative abundances are defined as $[X/Y] = \log( N(X)/ N(Y)) - \log( N(X)_\odot /N(Y)_\odot ) $, where $N(X)_\odot$ and $N(Y)_\odot$ are the reference solar abundances from \citet{Asplund09},  using the recommendations of \cite{Lodders09}.

%
 
    \begin{figure}
   \centering
   \includegraphics[width=9cm]{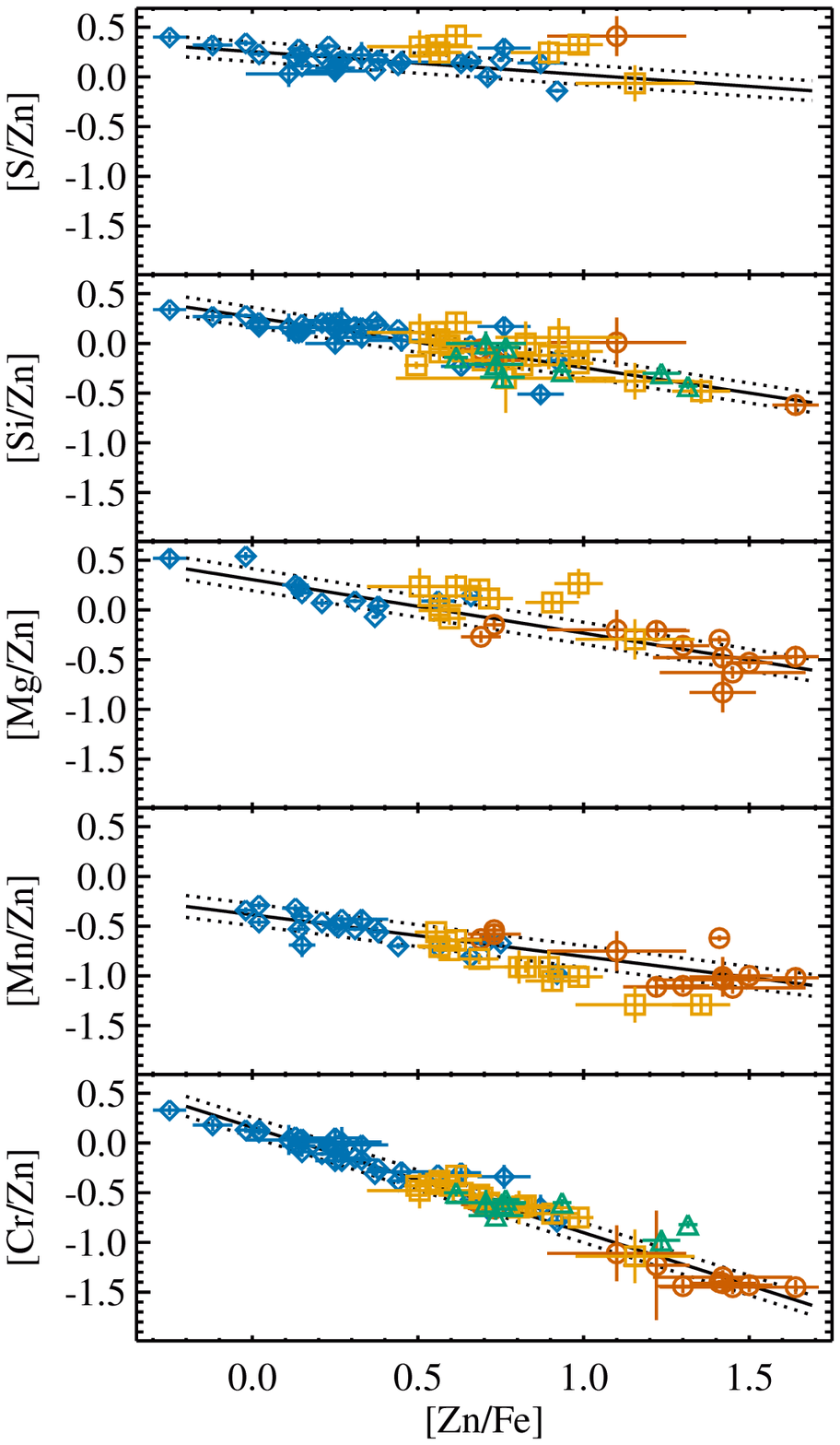}
   \caption{Relative abundances observed in the Galaxy \citep[orange circles,][]{Jenkins09}, SMC \citep[yellow squares,][]{Tchernyshyov15,Jenkins17}, LMC \citep[green triangles,][]{Tchernyshyov15}, and DLAs \citep[blue diamonds,][]{DeCia16}. The [Zn/Fe] is a proxy for the dust content, while the [$X$/Zn] traces the dust depletion of element $X$. The solid lines are from \citet{DeCia16} and were derived from a linear fit to the Galactic and DLA data, not including the Magellanic Clouds, where the intrinsic scatter is marked by the dotted lines.}
              \label{fig ZnFe}%
    \end{figure}
        \begin{figure}
   \centering
   \includegraphics[width=9cm]{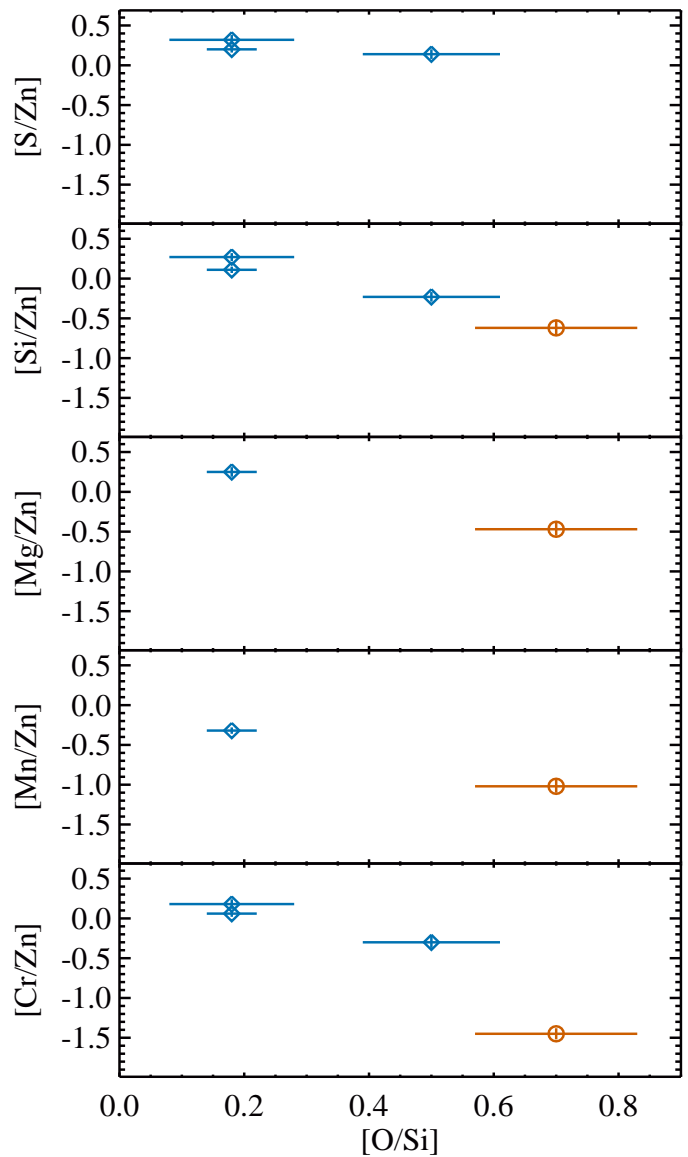}
   \caption{Same as Fig. \ref{fig ZnFe}, but using [O/Si] as a proxy for the dust content.}
              \label{fig OSi}%
    \end{figure}  
 
\section{ISM homogeneity}
\label{sec ISM}

In Fig. \ref{fig ZnFe} we directly compare relative abundances observed in the Galaxy, Small Magellanic Cloud (SMC), Large Magellanic Cloud (LMC), and DLAs. The data are taken from \citet{Jenkins09} for the Galaxy, \citet{Tchernyshyov15} and \citet{Jenkins17} for the SMC, \citet{Tchernyshyov15} for the LMC, and \citet{DeCia16} for DLAs. While metal abundances [$X$/H] cannot be directly compared  among different environments  because of the different metallicities of the systems, a meaningful comparison can be made for the relative abundances [$X/Y$]. 

The $x$-axis in Fig. \ref{fig ZnFe} shows the observed [Zn/Fe], which is a tracer of the amount of dust in the gas of each system. The reliability of [Zn/Fe] as a tracer of dust is discussed in Sect. \ref{sec znfe}. The $y$-axis shows the relative abundances of several metals, mostly refractory, with respect to Zn. Because Zn is only marginally depleted into dust grains, the relative abundances [$X$/Zn] are dominated by the dust depletion of the element $X$. In addition, [$X$/Zn] also carries the information of nucleosynthesis over- and underabundances with respect to Zn,  for example $\alpha$-element enhancement of S, Si, and Mg. The values of [$X$/Zn] at [Zn/Fe] $ = 0$ (i.e. when dust depletion is negligible) are indeed the nucleosynthesis over- or underabundances with respect to Zn, which are observed only at low metallicities and converge to zero at higher metallicities. The effects of nucleosynthesis are typically much smaller (up to $\sim0.3$~dex, and typically only for systems with low [Zn/Fe]) compared to the strong effects that dust depletion can have, up to two orders of magnitudes. The solid (and dotted) curves in Fig. \ref{fig ZnFe} show the empirical fit to (and intrinsic scatter of) the sequences of relative abundances from \citet{DeCia16}. These values were obtained from the DLA and Galactic data presented here, and already showed a similar behaviour to the Galactic ISM and DLA chemical properties. In this paper we include data for the Magellanic Clouds, to test whether the previously found sequences of relative abundances are valid for these systems as well.

A simple comparison of the observed relative abundances in DLAs, Magellanic Clouds, and the Galaxy shows that the same sequences of relative abundances are valid for all these environments. One exception is a small deviation for Mn in the SMC, Mn depleting more rapidly in the SMC than in the Galaxy. This has also been found by \citet{Jenkins17} via an independent analysis of the Mn depletion in the SMC. These authors explain the stronger depletion of Mn in the SMC as being due to a higher rate of Mn condensation in dust for the SMC, possibly because of the smaller amount of C-rich dust with respect to the Galaxy. 

The neutral ISM shows homogeneous chemical properties among very different environments (DLAs, Magellanic Clouds, and the Galaxy) at least in terms of chemical abundances, extending smoothly from systems with little  to no dust, to very dusty systems. This `universality' observed so far in the sequences of relative abundances has implications on the nature of DLAs and the origin of dust. 

First, the DLA gas behaves like the neutral ISM, at least from a chemical point of view. Strong shocks or heating in the circumgalactic medium (CGM) would probably disturb the chemical properties (e.g. by destroying the dust grains) and prevent the observed relative abundances from lining up on the sequences of Fig. \ref{fig ZnFe}. Therefore, the DLA gas seems most likely associated with the neutral ISM of high-$z$ galaxies. Searching for DLA galaxy counterparts in emission is challenging and has often had a low success rate \citep[e.g.][]{Krogager17}, but given the DLA chemical properties in absorption, it is unlikely that this gas is located in the CGM at very large distances from the DLA galaxy. This strengthens the association of DLAs with galaxies, as well as the motivation for using DLAs in the study of galaxies and their evolution.

Second, a dominant fraction of the dust budget is likely grown in the ISM, because the same sequences of relative abundances are valid for environments that have very different star formation histories and nucleosynthesis budgets.  The slopes of the sequences of relative abundances are roughly correlated with the condensation temperature of the different metals \citep{DeCia16}, implying that they are indeed driven by the presence of dust in the ISM. In mass, the dust in the ISM constitutes up to half of all the metals in the ISM, as shown by the values of the dust-to-metal ratios in the Local Group and extragalactic sources \citep{Watson11,DeCia13,Mattsson14}. Stellar dust production and destruction may take place and depend strongly on the star formation history of a galaxy. However, any contribution from stellar dust is limited or regulates in a way that does not interfere with the observed abundances in the ISM. While the seeds of dust may have different origins, the majority of dust mass seems to likely build up through grain growth in the ISM if the physical conditions are  favourable, such as high enough densities, metal content, and low enough temperatures. 

The relative abundances in the ISM reflect the effects of chemical evolution and metal reprocessing over the long evolution time of a galaxy. After a long enough time, we can expect similarities in the overall relative amounts of metals injected in the ISM, even for systems with different star formation histories. But while studying the ISM metals in local galaxies does not provide information on their stellar nucleosynthesis, it can be a powerful way to understand the properties of dust in the ISM.

\section{[Zn/Fe] as a tracer of dust in the ISM} 
 \label{sec znfe}

The observed [Zn/Fe] in the gas-phase can be used as a tracer of the dust content in the ISM. The reasons for this are that \textit{i)} Zn and Fe have very different refractory properties, Fe being heavily depleted, while Zn only mildly, and \textit{ii)} Zn and Fe tend to follow each other nucleosynthetically, or at least this is the net effect observed in the ISM. The second point has  recently been put into question \citep[e.g.][]{Berg15,Skuladottir17,Skuladottir18} because Zn is not an iron-peak element and has a complex nucleosynthetic origin. In particular, [Zn/Fe] has been observed to vary dramatically (down to $-0.8$~dex) in the stellar abundances of gas-poor environments such as dwarf spheroidals and the Galactic Bulge \citep[e.g.][]{Skuladottir17,Duffau17,Skuladottir18}. In addition, strong deviations (up to $+0.5$~dex) of the stellar [Zn/Fe] are observed at  low ($[\mbox{Fe/H}]<-2$) and high metallicity ($[\mbox{Fe/H}]>0$) in the Galaxy \citep[e.g.][]{Primas00,Cayrel04,Nissen07}. In particular, for metallicities [Fe/H] $<-2.0$, Zn starts behaving like an $\alpha$ element, although with a small amplitude, and therefore is no longer a robust proxy for Fe in stellar abundances. This variability of [Zn/Fe] in some stellar environments may lead the reader to cast some doubts on the use of [Zn/Fe] as a zero-reference for the study of the ISM,  and indeed the intrinsic properties of [Zn/Fe] in the ISM need to be carefully assessed before we can safely use them to characterize the dust content in the ISM.

However, the values of the [Zn/Fe] in the ISM are by far more homogeneous than in stellar abundances, i.e. free from strong nucleosynthetic deviations, and especially in the metallicity range which is most relevant for DLAs, for the following reasons:\\  
\textit{i)} The stellar values of [Zn/Fe] is flat and $\sim0$ in the Galaxy, in the metallicity range $-2\lesssim [\mbox{Fe/H}]\lesssim 0$ \citep[e.g.][]{Sneden91,Matteucci93}, which is the most relevant for DLAs;\\
\textit{ii)} The [Zn/Fe] observed in the ISM in the Galaxy and DLAs correlates with the strength of the dust depletion $F*$ \citep{Jenkins09,DeCia16};\\
\textit{iii)} No strongly negative [Zn/Fe] are observed in the ISM of DLAs, even considering that the effects of dust depletion at low [Zn/Fe] are basically negligible. Negative [Zn/Fe] are very rare, the minimum observed in DLAs being is $\sim -0.2$~dex. It may in fact be possible that the reference [Zn/Fe] is lower than zero, down to $-0.2$~dex, due to some small deviations of Zn nucleosynthesis with respect to Fe, and this could in principle shift  the origin of the depletion sequences a bit. However, this cannot be robustly constrained yet because of the paucity of data;\\
\textit{iv)} We observed tight correlations (scatter $\lesssim -0.2$~dex) between [Zn/Fe] and [$X$/Zn] (Fig. \ref{fig ZnFe}). These can exist only if [Zn/Fe] is not strongly affected by nucleosynthesis deviations;\\
\textit{v)} The same tight correlations are also observed in the Galaxy and DLAs between [Zn/Fe] and [$X$/S], and between [Zn/Fe] and [$X$/P] \citep[Fig. 3 of][]{DeCia16};\\
\textit{vi)} The [Zn/Fe] observed in the ISM in DLAs correlates with [Ti/Si] \citep[Fig. A.1 of][]{DeCia16} and other minor dust tracers, such as [O/Si], which are not affected by nucleosynthesis effects (Ti, Si, and O are all $\alpha$ elements);\\
\textit{vii)} The sequences of relative abundances are still visible even when using [O/Si] instead of [Zn/Fe] as a dust tracer, which is not affected by nucleosynthesis effects. This is shown in Fig. \ref{fig OSi}. However, [O/Si] is a worse dust tracer than [Zn/Fe] because the range of [O/Si] has a much smaller amplitude,  about half of the [Zn/Fe] range, due to the lesser difference in refractory properties between O and Si. In addition, only limited data is currently available with sufficient metal observations to populate  these sequences well.\\  

Therefore, we conclude that [Zn/Fe] can be considered as a reliable dust tracer in the ISM, at least in the metallicity range $-2\lesssim [\mbox{Fe/H}]\lesssim 0$, which is typical of most DLAs. Any potential nucleosynthesis deviation of [Zn/Fe] from zero must be limited, and smaller than $\sim0.2$~dex. Further investigations are needed to better quantify this in the future. Despite the intrinsic variations of [Zn/Fe] in stars in low-metallicity or gas-free environments, the intrinsic (dust-free) [Zn/Fe] in the ISM is almost constant and consistent with zero. This is perhaps the result of metal reprocessing in the ISM, where the contribution of different stellar populations is mixed in time. Zinc has a complex nucleosynthetic origin, being produced in both explosive nucleosynthesis and $s$-processes \citep[e.g.][]{Matteucci93}. In a way, Zn can be considered an $\alpha$-element with a very low amplitude in some environments \citep[e.g.][]{Duffau17},  but given the observations of the ISM relative abundances, these differences are washed away in time by metal recycling in the ISM.

%
%
 
\section{Conclusions}
\label{sec conclusions}

We compared the chemical properties of the ISM in the Galaxy, DLAs, and for the first time we included the Magellanic Clouds as well. We observed sequences of relative abundances that are driven by the presence of dust in the ISM, and are valid for the Galaxy, Magellanic Clouds, and DLAs. This shows homogeneous properties of the neutral ISM in different environments and $z$, at least regarding the metal and dust content. The `universality' of the sequences of relative abundances observed so far leads  us to further conclude that a dominant fraction of the dust budget is likely built up through grain growth in the ISM if the physical conditions of the gas are favourable, and regardless of the star formation history of the system. While the contribution of stellar dust production and destruction can still be present, it must be limited or regulated to maintain the observed chemical properties in the ISM in different environments. We recommend taking this into account in models of dust formation or chemical evolution.

In addition, we conclude that the DLA gas mostly behaves like the neutral ISM, at least from a chemical point of view, and is unlikely to be associated with the CGM at large distances from the DLA galaxy. This strengthens the case for using DLAs as a complementary tool to study galaxies, out to high redshift and down to low masses. Finally, we demonstrate the validity of [Zn/Fe] as a robust dust tracer in the ISM of different environments, from the Galaxy to DLAs.

\begin{acknowledgements}
    ADC thanks Prof. Edward Jenkins, Prof. Francesca Matteucci, and C\'edric Ledoux for the insightful discussions, and the anonymous referee for the useful comments. 
\end{acknowledgements}

  \bibliographystyle{aa} 
  
 \bibliography{biblio.bib} 

\end{document}